\documentclass{sig-alternate-per}

\usepackage{tikz}
\usepackage{todonotes}
\usepackage{xcolor}
\usepackage{url}
\usepackage{csquotes}
\usepackage{censor}
\usepackage[frozencache,cachedir=.]{minted}
\usepackage{soul}

\usepackage{url}

\usepackage{breakurl}
\usepackage[breaklinks,colorlinks=true,allcolors=cyan,citecolor=magenta]{hyperref}

\title{Moving Beyond \st{Marginal Carbon Intensity}}
\subtitle{\vspace{-0.5em}A Poor Metric for Both Carbon Accounting and Grid Flexibility\vspace{0.5em}}

\usepackage{titlesec}
\titlespacing{\section}{0pt}{*2}{*1}

\author{
  Philipp Wiesner\vspace{1mm}\\
  \large\sf Technische Universität Berlin\\
  \large\sf wiesner@tu-berlin.de
  \and
  Odej Kao\vspace{1mm}\\
  \large\sf Technische Universität Berlin\\
  \large\sf odej.kao@tu-berlin.de
}

\conf{CarbonMetrics 2025}
\confinfo{Stony Brook, NY}
\toappear{Workshop on Measurements, Modeling, and Metrics for Carbon-Aware Computing (CarbonMetrics) @ ACM SIGMETRICS '25, \the\confinfo\par \the\copyrightetc.}

\begin{document}

\maketitle

\begin{abstract}
\noindent
Marginal Carbon Intensity (MCI) has been promoted as an effective metric for carbon-aware computing.
Although it is already considered as impractical for carbon accounting purposes, many still view it as valuable when optimizing for grid flexibility by incentivizing electricity usage during curtailment periods.
In this statement paper, we argue that MCI is neither reliable nor actionable for either purpose. 
We outline its fundamental limitations, including non-observability, reliance on opaque predictive models, and the lack of verifiability.
Moreover, MCI fails to reflect curtailment caused by high-carbon sources and offers no insight into the quantity of available excess power.
We advocate moving beyond MCI and instead call for research on more actionable metrics, such as direct reporting of excess power, explicit modeling of energy storage and grid stability, and integration with emerging granular renewable energy certificate markets.
\end{abstract}


\section{Introduction}

Carbon-aware computing has emerged as a prominent research field focused on aligning the electricity demand of digital infrastructure with the availability of clean energy, typically by shifting computational workloads across geographically distributed data centers or deferring them in time.
However, \enquote{carbon-aware} currently serves as an umbrella term for two different optimization objectives.
Most existing approaches focus on \emph{carbon accounting}, aiming to reduce the operational carbon emissions of electricity consumption~\cite{Radovanovic_Google_2022, hanafy2024asplos, gsteiger2024caribou, carbonscaler2024sigmetrics, sukprasert2024eurosys, wiesner2025qualitytime, murillo2024cdn_shifter}, in other words, the reported Scope~2 emissions under the GHG Protocol~\cite{ghg_protocol}.
In contrast, a large body of work acknowledges that today's methodologies for carbon accounting fail to capture the complexities of electric grids~\cite{Schneider2019Double, Bjoern2022} and optimize for \emph{grid flexibility}~\cite{Lin_CouplingDatacentersPowerGrids_2021, Zheng_MitigatingCurtailment_2020, dodge2022carbon_intensity_ai, lindberg2022geographic_shifting, wiesner2024fedzero, Wiesner_Cucumber_2022, SEPIA_NegotiationGameITEnergyManagement_2020}: demand side management to address system-level inefficiencies such as grid congestion, storage constraints, and renewable energy curtailment.

The two objectives typically rely on different metrics to guide optimization.
In the context of carbon accounting, average carbon intensity (ACI) has become the predominant metric. 
ACI represents the production-weighted average of emission factors from all power sources within a given region.
Emission factors describe the amount of carbon emitted per kWh of generated electricity of an energy source.
For example, the IPCC Fifth Assessment Report~\cite{IPCC_Annex3_2014} reports a life-cycle emission factor of 820 gCO\textsubscript{2}/kWh for coal, but only 48 gCO\textsubscript{2}/kWh for utility-scale solar energy.
ACI offers a transparent and verifiable methodology that supports consistency and reproducibility in carbon accounting. 
As it is mentioned in the GHG protocol~\cite{ghg_protocol}, ACI is likely to play an important role in future regulations, an assumption widely adopted in recent works on carbon-aware optimization~\cite{gsteiger2024caribou, hanafy2024asplos, sukprasert2024eurosys, Radovanovic_Google_2022}.

Marginal carbon intensity (MCI)\footnote{\small Terminology varies: MCI is also referred to as the Marginal Operating Emissions Rate~(MOER)~\cite{wattime2022moer}, the Locational Marginal Carbon Emissions~(LMCE)~\cite{gorka2025electricityemissions}, or the Marginal Emission Factor~(MEF)~\cite{koebrich2025evaluation} in related works.} offers an alternative approach by estimating the \emph{additional} emissions resulting from incremental changes in electricity demand. 
Rather than averaging across all sources, MCI represents the emission factor of the marginal generator---the power plant that adjusts its output to meet demand changes. 
While ACI averages emissions over relatively large geographic areas and time frames, MCI establishes a direct cause-effect relationship for load shifting: when fossil-fueled generators are on the margin, MCI is high and incentivizes demand reductions; when renewables are on the margin (typically during oversupply) MCI drops toward zero, encouraging consumption when clean energy would otherwise be curtailed.

In theory, this makes MCI a compelling metric for demand management, which is why organizations like Microsoft and WattTime have advocated for its use~\cite{buchanan2023carbonaware_computing, watttime2017locationlocationlocation}.
Consequently, recent works have explicitly leveraged MCI in carbon-aware scheduling frameworks and analyses~\cite{sukprasert2024average_vs_marginal, dodge2022carbon_intensity_ai, lindberg2022geographic_shifting, lindberg2021eenergy}.
However, in this statement paper, we argue that more reliable and actionable metrics than MCI exist---both for carbon accounting and optimizing grid flexibility.

\section{MCI is Impractical for Carbon Accounting}

The practical implementation of MCI faces major challenges, which have been extensively discussed by Electricity Maps, a leading provider of carbon intensity data, in a series of technical blog posts~\cite{electricitymapsMarginal_4, electricitymapsMarginal_5, electricitymapsMarginal_6, electricitymapsMarginal_7}.
In summary, the fundamental issue with MCI is that it cannot be measured or derived from other measurable metrics.
Unlike ACI, which is directly derived from public electricity data, MCI attempts to capture the emissions impact of a marginal increase in electricity demand. 
However, electricity dispatch is not a simple, deterministic process. It is shaped by market dynamics, transmission constraints, and decentralized decision-making, making it virtually impossible to pinpoint which generator responds to a small change in demand.
As a result, current methodologies rely on predictive models trained on historical grid behavior and market prices. These models are inherently probabilistic and often opaque, introducing significant uncertainty and undermining the transparency required for regulatory or compliance use. WattTime, a major provider of MCI estimates, explicitly acknowledges this limitation:
\enquote{Marginal emissions are not directly measurable [...] Without ground truth, it can be challenging to determine which models are closer to the truth and quantify their accuracy.}~\cite{watttime_methodology}.
This lack of verifiability is particularly problematic for carbon accounting, since different modeling choices can lead to drastically different MCI estimates.

Due to these issues---non-observability, probabilistic estimates, model dependance, and lack of verifiability---Elec\-tric\-i\-ty Maps has recently discontinued support for MCI~\cite{electricitymaps2024marginal}.
However, they acknowledge that \enquote{marginal emission factors may well have a future} outside of Scope~2 accounting.
In the following, we argue that also for optimizing grid flexibility, better metrics exist.

\section{Excess Power is a More Actionable Metric for Optimizing Grid Flexibility}

Unfortunately, today's carbon accounting methodologies provide an incomplete and distorted picture of electricity-related emissions and fail to address electricity curtailment and grid stability.
A main reason for this is a lack of high-resolution monitoring data from power grids, both in time and space.
Yet, any marginal metric inherently depends on such granular data, implying that their widespread adoption assumes future improvements in grid observability.
\emph{Given that such data becomes available}, we argue that already simple metrics like directly quantifying the amount of excess renewable power during curtailment periods---as already done in many related works~\cite{journal2017geo_load, SEPIA_GreenITscheduling_2018, SEPIA_NegotiationGameITEnergyManagement_2020, Zheng_MitigatingCurtailment_2020, Wiesner_Cucumber_2022, wiesner2024fedzero}---offers a significantly more actionable signal for carbon-aware optimization than MCI.
To illustrate this, we construct a toy example based on a grid with only three power sources:
\begin{enumerate}
	\item \textbf{Solar}: A low-carbon source with variable output that cannot be controlled directly.
	\item \textbf{Coal}: A high-carbon source that is used to meet base\-load and adjusts only slowly to demand fluctuations.
	\item \textbf{Gas}: A moderately carbon-intensive but highly flexible source that can be adjusted in real time to meet residual demand.
\end{enumerate}

\begin{figure}
    \centering
    \includegraphics[width=0.95\columnwidth]{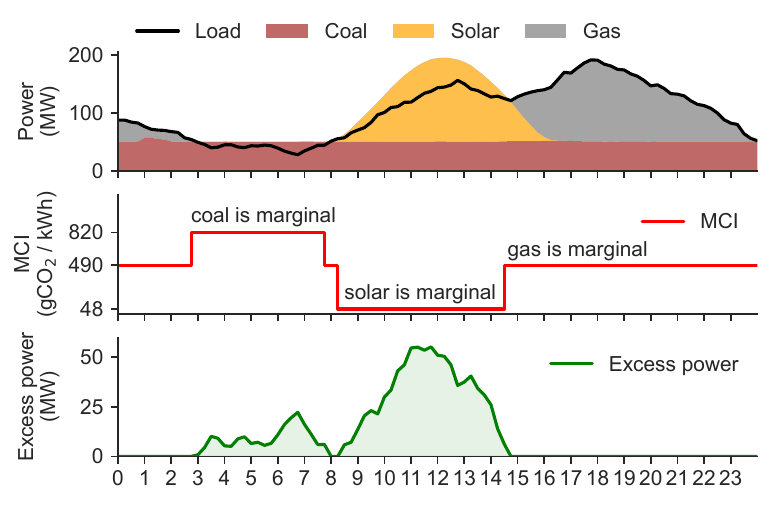}
    \caption{Comparison of MCI and \emph{excess power}. Excess power also reflects curtailment periods caused by high-carbon sources and informs about the amount of curtailed energy.}
    \label{fig:mci_vs_ep}
\end{figure}

Figure~\ref{fig:mci_vs_ep} depicts exemplary power production and demand during one day, alongside the corresponding MCI and excess power.
Note that, for MCI, low values incentivize electricity usage, while for excess power, high values incentivize electricity usage.
Both metrics encourage energy usage between 08:10 and 14:30, where excess solar power is available.
However, excess power has two key advantages over MCI.

First, excess power accounts for curtailments from any energy source, not just low-carbon sources.
For example, coal power plants are highly inflexible and require several hours to adjust their output.
As a result, any mismatch between forecasted and actual demand can lead to excess power, even in the absence of renewables.
In our example, between 03:00 and 08:00, MCI reports a high value, discouraging consumption, despite excess power being available.
We argue that \emph{from a consumers's perspective, the identity of the marginal generator is irrelevant}.
As a consumer, the operational details behind excess power lie outside your control---what matters is whether additional consumption can be met without increasing fossil-based generation.

Second, excess power provides a quantitative measure of how much additional energy can be consumed at a given time without increasing emissions.
This prevents overconsumption, which could trigger the ramp-up of more carbon-intensive power sources.
In contrast, MCI only estimates the carbon intensity of the current marginal generator but offers no insight into when the grid will switch to a new marginal source.
As a result, it lacks the necessary information to make precise carbon-aware scheduling decisions.

In practice, the usefulness of excess power as a metric depends on the availability of curtailment data from grid operators. 
A few regions already offer this: California's CAISO, for instance, publishes hourly curtailment volumes by source and cause; similarly, Great Britain, Australia, and parts of continental Europe report granular redispatch and spillage data.
Others, do not publish such data directly but make it possible to infer excess power from dispatch and telemetry feeds. 
Since most grids already track generator setpoints and actual output, it is technically feasible for third-party providers to estimate excess power signals, even in regions where system operators do not yet report them.

\section{Toward Actionable Marginal Metrics}

We conclude that MCI, despite its conceptual appeal, does neither capture the needs of carbon-aware approaches focused on reducing Scope~2 emissions (a clear, transparent, and verifiable methodology) nor of efforts to optimize grid efficiency.
Rather than investing further effort into identifying the elusive marginal generator, which is neither practically verifiable nor particularly helpful for demand management, research focus should shift toward more actionable marginal signals.
In addition to directly predicting curtailment periods and available excess power, such metrics could consider:

\begin{enumerate}
    \item \textbf{Energy storage}: Batteries can hide real-time grid conditions by shifting demand without indicating if excess power is actually available. To ensure meaningful flexibility, marginal metrics should account for round-trip losses, degradation, and state of charge, and penalize actions that merely cycle batteries without relieving grid stress.

    \item \textbf{Market indicators}: In regions with locational mar\-gi\-nal pricing, low/negative prices can already serve as a proxy for curtailment~\cite{acun2023curtailmentprediction}. In contrast, in Europe’s zonal and heavily regulated markets, it currently makes little sense to base marginal metrics on prices, as they poorly reflect local grid conditions.
    	However, recent proposals for time- and location-granular renewable energy certificate markets aim to create more accurate, standardized signals~\cite{Google_CarbonCounting_2021,Microsoft_CarbonCounting_2021}.
        If adopted, these could support carbon-aware scheduling by aligning emissions impact with financial incentives.
    
    \item \textbf{Grid stability}: The steep, synchronized power ramps of modern GPU clusters can impose significant stress on both utility grids and on-site microgrids~\cite{patel2024characterizing}. For example, Meta notes that \enquote{During training, tens of thousands of GPUs may increase or decrease power consumption at the same time [...]. When this happens, it can result in instant fluctuations of power consumption across the datacenter on the order of tens of megawatts}~\cite{grattafiori2024llama}. Metrics should account for such dynamic load behavior and incentivize smoother load profiles to reduce strain on grid infrastructure.
        
    \item \textbf{Trade-offs}: 
		Carbon-aware strategies can involve trade-offs across multiple levels. 
		At the system level, some measures that improve carbon efficiency can harm energy efficiency~\cite{hanafy2023war}.
		At the service and user level, shifting or degrading workloads may interfere with quality of experience and trigger rebound effects~\cite{wiesner2025qualitytime}.
		Future metrics should aim to make these trade-offs explicit.
\end{enumerate}

There is a clear need for marginal metrics, as ACI obscures the operational realities of modern power systems and electricity markets. 
This shortcoming is one of the main drivers behind the ongoing revision of Scope 2 accounting in the GHG protocol~\cite{ghg_scope2_twg}, and highlights that the debate around marginal signals is far from over. 
However, if we aim to capture the complexity of the grid in a single actionable metric, we can---and should---do better than MCI. 

\bibliographystyle{abbrv}
\bibliography{bib}

\end{document}